\documentclass[conference]{IEEEtran}
\usepackage{graphicx,amsmath,epsfig,times}
\usepackage{psfrag,cite}
\usepackage{amstext,amssymb,latexsym}
\usepackage{setspace}

\DeclareMathOperator*{\argmax}{arg\,max}

\begin{document}

\title{Practical Wireless Network Coding and Decoding Methods for Multiple Unicast Transmissions}
\author{\IEEEauthorblockN{Tu\u{g}can~Akta\c{s}}
\IEEEauthorblockA{Dept. of Electrical and Electronics Eng.,\\
Middle East Technical University,\\
Ankara, Turkey\\
e-mail: taktas@metu.edu.tr}
\and
\IEEEauthorblockN{Ali~\"{O}zg\"{u}r~Y\i lmaz}
\IEEEauthorblockA{Dept. of Electrical and Electronics Eng.,\\
Middle East Technical University,\\
Ankara, Turkey\\
e-mail: aoyilmaz@metu.edu.tr}
\and
\IEEEauthorblockN{Emre~Akta\c{s}}
\IEEEauthorblockA{Dept. of Electrical and Electronics Eng.,\\
Hacettepe University,\\
Ankara, Turkey\\
e-mail: aktas@ee.hacettepe.edu.tr}
}

\maketitle
\begin{abstract}
We propose a simple yet effective wireless network coding and decoding technique. It utilizes spatial diversity through cooperation between nodes which carry out distributed encoding operations dictated by generator matrices of linear block codes. For this purpose, we make use of greedy codes over the binary field and show that desired diversity orders can be flexibly assigned to nodes in a multiple unicast network, contrary to the previous findings in the literature. Furthermore, we present the optimal detection rule for the given model that accounts for intermediate node errors and suggest a network decoder using the sum-product algorithm. The proposed sum-product detector exhibits near optimal performance.
\end{abstract}


\section{Introduction}

In order to counteract the effects of fading in wireless communication networks, many ways of creating diversity for transmitted data have been proposed. Utilizing the spatial diversity inherent in wireless channels, cooperative communication \cite{Erkip03} has been of great interest in recent years. In \cite{Laneman04, Laneman06} three methods to be used by relay nodes are described: amplify-forward (AF), decode-forward (DF) and demodulate-forward (DMF). The AF method attains full diversity, whereas other two cannot, unless the propagation of errors resulting from the decoding operation is avoided. One of the various ways to handle this problem is using CRC-based methods, which results in loss of spectral efficiency due to drop of a packet with only a few bit errors. An on/off weighting based on relay signal-to-noise power ratio (SNR) is given in \cite{Onat08}. Weighting of the signals either at the relay or at the receiver using the relay error probability is proposed in \cite{Wang07, Wang08}. Yet another idea is transmitting the log-likelihood ratios (LLR) of bits \cite{Yang07}. However, the quantization or peak-to-average power ratio problems are inherent for these methods. In addition, both the AF and DF methods lead to high complexity especially for the relays. As an alternative, relays may use the simple DMF method, which is shown to avoid error propagation in \cite{Laneman06}, if the error probabilities at relays are known and the maximum a posteriori probability (MAP) detection is employed at the receiver. In this paper, we will concentrate on MAP-based detection rules at the receiver and DMF at the intermediate (relay) nodes due to ease of implementation.

Network coding (NC) was initially proposed to enhance network throughput in wired systems with error-free links of unit capacity\cite{Ahlswede00}. Later studies exhibited the good performance of random linear NC \cite{Ho06}. In wireless networks (nodes naturally overhearing transmissions), NC can be utilized to create diversity, reduce routing overhead, and introduce MAC layer gains as discussed for practical systems in \cite{Medard08}. Although most of the work in the literature concentrate on the multicast transmission \cite{Renzo10}; we deal with a subset of network involving multiple unicast transmissions, which is inherent in real-life scenarios. Improving diversity orders of data symbols via spatial opportunities (multipath diversity) is our motivation. We formulate the multiple unicast transmission problem such that for each unicast transmission, there is a distinct diversity order. The major goal in this paper is to introduce \textit{practical NC/decoding methods} for improving the diversity order of a network with the overall rate of transmission in mind. 

We consider a simple NC scheme based on DMF. Given a relay combining strategy, which we represent by a generator matrix and a vector of transmit schedule, we investigate the diversity order of each source, which can be unequal. We propose a novel method for designing the generator matrix based on greedy codes over the binary field. The proposed method is very flexible in that any desired level of diversities of the sources can be achieved with the highest network coding rate possible. The analysis relies on the fact that optimal MAP detector which the employs reliability information of the relays, avoids loss of diversity due to error propagation. The numerical complexity of the MAP detector can be impractical. Thus we propose a practical close approximation of the MAP detector: the sum-product detector. 
 
A study based on flexible network codes in a two-source two-relay system with emphasis on unequal error protection is \cite{Renzo11}, where authors propose a suboptimal detection rule (distributed minimum distance detector) that is known to result in diversity order loss. Note that our model is more general, and captures full diversity due to use of sum-product detector with relay reliability information. One of the studies closest to ours is \cite{Xiao10}, where the NC operation is fixed in construction yielding very large Galois Field (GF) sizes for increasing network size and relay nodes carry out complex DF operation for each transmission they overhear. However, our results indicate that any diversity order can be achieved for any unicast transmission even with GF of size 2, using greedy codes and simple DMF operation.
\section{Wireless Network Model}
\label{sec:wn_model}
\subsection{Demodulate and Forward Wireless Network}
\label{sec:wn_model_demod}
In this work, we analyze a wireless network in which unicast transmission of data symbols, each belonging to a different source, is to be carried out utilizing NC at the intermediate nodes. Under the general operation scheme, every node may act both as a member (source or destination) of a unicast communication pair and as an intermediator (relay) node for other unicast pairs. Consider a subset of nodes in which there are $k$ nodes transmitting data to a single receiver node, and every transmission is heard by every other node. Let the symbol transmitted by node $i$ be denoted by $u_i$, for $i \in \{ 1, \dots, k\}$. We assume $u_i$ to be statistically independent. The receiver is the destination for one or more of the source symbols, and acts as a relay for the others. The receiver may try to detect the data symbols for which it is the destination with a higher priority. In such cases, a transmission strategy which provides unequal error protection can be useful.
  
The transport of $k$ symbols are realized over $n$ transmissions, which form a \textit{round} of network coded communication (see Fig. \ref{fig:network3} for a simple network with $k=3$ and $n=4$). We assume that these transmissions are done in orthogonal channels so that the strict synchronization requirements between transmitting nodes are relaxed and the complexity is not increased due to the interference cancellation techniques at the receiving nodes. The channel may be shared by a time division multiple access technique for simplicity in model description.

Let $\mathbf{u}=[u_1 \ u_2 \ \ldots \ u_k]$ be the combined data vector for $k$ source nodes in the subnetwork, where $u_i$ is an element from the Galois field of size $M$, GF($M$). In time slot $j\in \left\lbrace 1,\ldots ,n\right\rbrace$, a transmitting node $v_j\in \left\lbrace 1,\ldots ,k\right\rbrace$ forms a linear combination of its own and other nodes' data. If $v_j$ has detected all data to be encoded correctly, it simply forms $c_j = \mathbf{u}\mathbf{g}_j$, where $\mathbf{g}_j$ is a $k \times 1$ network encoding vector whose entries are elements of GF($M$). In case $v_j$ has detected at least one of the data of nodes $\left\lbrace 1,\ldots ,k\right\rbrace \setminus v_j$ incorrectly, i.e. $\hat{u}_i \neq u_i$ for some $i\in \left\lbrace 1,\ldots ,k\right\rbrace \setminus v_j$, it forms $\hat{c}_j= \mathbf{\hat{u}}\mathbf{g}_j$ that is also an element of GF($M$). Then $v_j$ modulates and transmits this symbol as $s_j=\mu(\hat{c}_j)$ to the receiver node $0$: 
\begin{equation}
\label{eqn:coded}
s_j=\mu(\hat{c}_j)=\mu(\mathbf{\hat{u}} \mathbf{g}_j),
\end{equation}
where $\mu(.)$ is the mapping of a coded symbol to a constellation point. Although symbols may come from any alphabet and non-binary constellations may be used, we will focus here on GF($2$) and binary phase-shift keying (BPSK), which means $s_j = 1-2c_j$. Our assumption is that each vector $\mathbf{g_j}$, source address $v_j$ and probability of error $p_{e_j}$ for the transmitted symbol are appended to the corresponding packet and are known at the receiving nodes. We work on transmissions with no channel coding and deal with single network coded data symbol $c_j$ as a representative of symbols within a packet transmitted by node $v_j$. Hence there is only one index (the transmission time slot) in $c_j$. At the end of a round of transmissions, if no errors occur at the intermediate nodes, the overall vector of $n$ symbols coded cooperatively in the network can be written as
\begin{equation}
\mathbf{c}= \left[c_1 c_2 \hdots c_n \right] =\mathbf{u} \left[ \mathbf{g}_1 \ \mathbf{g}_2 \ \hdots \ \mathbf{g}_n \right]=\mathbf{u} \mathbf{G}, 
\end{equation}
where $\mathbf{G}$ is the generator matrix (named as transfer matrix in \cite{Xiao10}). The vector of transmitting nodes is denoted by 
\begin{equation}
\mathbf{v}= \left[ v_1 v_2 \ldots v_n\right].
\end{equation}
The choices $\mathbf{u}, \mathbf{G}, k, n$ for the parameters defining the network are not arbitrary. They are used intentionally to point out the analogy to regular linear block codes. However, reliable detection of all data symbols, i.e., whole block $\mathbf{u}$, originating from a single error-free source is of interest for a regular decoder; whereas node $0$ may desire to reliably detect, as an example, only $u_1$ using $\mathbf{c}$. This difference and the diversity orders of distinct symbols are clarified in Section \ref{sec:nc_const}.
\subsection{Optimal Network Decoding Using Reliability Information}
\label{sec:wn_model_optimal}
The intermediate nodes are assumed to use the demodulate and forward technique due to its simplicity. In a wireless network, an intermediate node $v_j$ has a noisy detection result $\hat{\mathbf{u}}$ of $\mathbf{u}$. Thus, (\ref{eqn:coded}) can be rewritten as 
\begin{equation}
\label{eqn:noisy2}
s_j=\mu(\hat{c}_j)=\mu(c_j+e_j),
\end{equation}
where $e_j$ denotes this propagated error and we observe that a possible error in $\hat{\mathbf{u}}$ propagates to $\hat{c}_j$ after the network encoding operation dictated by $\mathbf{g}_j$ is realized. We assume that node $v_j$ knows the probability mass function of $e_j$, or equivalently the relay reliability information. 
 The received signal by node $0$ at time slot $j$ is then $y_j=h_js_j+w_j$,
where $h_j$ is the channel gain coefficient resulting from fading during the $j$th time slot for the link between node $v_j$ and node $0$ and $w_j$ is the noise term for the same link. The fading coefficient is circularly symmetric complex Gaussian (CSCG), zero-mean with variance $E_s$, i.e., it has distribution $\mathbb{CN}(0,E_s)$. The noise term is CSCG with $\mathbb{CN}(0,N_0)$. The usual independence relations between related random variables representing fading and noise terms exist. The overall observation vector of length $n$ at node $0$ is
\begin{align}
\label{eqn:obs}
\mathbf{y}=\mathbf{H} \mathbf{s}+\mathbf{w},
\end{align}
where ${\mathbf{y}}=[y_1 \ \ldots y_n]^T, {\mathbf{s}}=[s_1 \ \ldots s_n]^T=\mu (\hat{\mathbf{c}}^T), {\mathbf{w}}=[w_1 \ \ldots w_n]^T$ and $\mathbf{H}$ is a diagonal matrix whose diagonal elements are independent channel gains $h_1,h_2,\ldots,h_n$. It is assumed that $\mathbf{H}$ is perfectly known at the receiver. Combining the coded symbols in a network code vector, we obtain
\begin{equation}
\label{eqn:codevect}
\mathbf{\hat{c}} = \mathbf{c} + \mathbf{e} = \mathbf{uG} + \mathbf{e},
\end{equation}
where $\mathbf{e} = \left[ e_1 \ldots e_n\right]$ is the error vector. We assume that $\mathbf{e}$ is independent of $\mathbf{c}$ although dependence can be incorporated in the detectors to be developed. 
As a result, using (\ref{eqn:noisy2}), (\ref{eqn:obs}), and  (\ref{eqn:codevect}), the observation vector at node $0$ is
\begin{equation}
\mathbf{y} = \mathbf{H} \ \mu(\mathbf{uG} + \mathbf{e})^T +\mathbf{w}.
\end{equation}
Thus node $0$ has access to the likelihood $p(\mathbf{y} \vert \mathbf{u,e})$ and $p(\mathbf{e})=\prod_{j=1}^{n}p(e_j)$, assuming the errors are independent. In order to avoid the propagation of errors occurring at intermediate nodes, node $0$ has to utilize the reliability information $p(\mathbf{e})$ as given in \cite{Laneman06}. Then, at the receiver, MAP estimate of the source bit of interest, say $u_1$ (denoted by $\hat{\hat{u}}_1$), can be obtained as:
\begin{align}
\label{eqn:opt_det}
\hat{\hat{u}}_1 = \argmax_{u_1}p(u_1\vert \mathbf{y})= \argmax_{u_1}\sum_{u_2,\ldots ,u_k}\sum_{\mathbf{e}} p(\mathbf{y} \vert \mathbf{u,e})p(\mathbf{e}),
\end{align}
which is the individually optimum detector for $u_1$. 
As a result, for the optimal detection of $u_1$, the receiver node needs the reliability information vector: $\mathbf{p_{e}} = [p_{e_{1}} \ldots p_{e_{n}}]$, where  $p_{e_{j}}$ depends on the probability mass function of $e_j$. We observe the performance of this detector in Section \ref{sec:num_results_sub1} assuming that instantaneous reliability value for each bit of codeword $\mathbf{\hat{c}}$ is appended to the packet by the intermediate node.  

The main problem related to the MAP-based detection rule of (\ref{eqn:opt_det}) is the complexity of required operations. Therefore we suggest a practical network decoding technique in Section \ref{sec:spnd}.

\section{Linear Block Codes Used as Network Codes}
\label{sec:usp}
When the conventional block coding is considered, the average error performance over all data symbols is of interest. Therefore, for a linear block code, the main metric utilized for comparison is the minimum distance\begin{footnote}{
Minimum distance is equal to the diversity order in the case that independent channels are used for transmission of coded symbols \cite{Proakis}.}\end{footnote}. However, there are distinct minimum distances (defined as \textit{separation vector} in \cite{Gils83}) for different data symbols, whenever we are interested in performance of individual symbols that may originate from different source nodes as with NC. This idea is exemplified in \cite{Renzo11} in the context of NC for simple networks. We will generalize and use this idea for investigating diversity orders assigned to source symbols in a network. For demonstration, let us start with a simple $M$-ary symmetric channel model for the transmission of each one of the $n$ symbols. Then the received vector at node $0$ is $\mathbf{r}=\mathbf{u} \mathbf{G} + \bf{t}$,
where $\mathbf{t}$ is the $1 \times n$ error vector of independent terms from GF($M$). 
For the case of conventional coding, the joint MAP decoding 
\begin{align}
\hat{\hat{\mathbf{u}}}=\argmax_{\mathbf{u}} p(\mathbf{u}\vert \mathbf{r}) =\argmax_{\mathbf{u}} p(\mathbf{r} \vert \mathbf{u})
\end{align}
is used, where all vectors of source data symbols are assumed to be equally likely. Therefore, an error is the event that at least one of the detected symbols $\hat{\hat{u}}_i$ is different than the original symbol $u_i$, i.e., $\hat{\hat{\mathbf{u}}} \neq \mathbf{u}$. In contrast, in NC, the priority of individual sources may happen to be different from the point of view of the receiver and erroneous detection of high-priority symbols may determine the performance figure. Here, the optimal way of detecting distinct symbols follows individual MAP decoding:
\begin{align}
\hat{\hat{u}}_i&=\argmax_{u_i} \sum_{\lbrace u_1,\ldots,u_k\rbrace \setminus u_i}p(\mathbf{r} \vert u_1,u_2,\ldots,u_k)
\end{align}
and we are interested in errors $\hat{\hat{u}}_i \neq u_i$, where priority may depend on $i$. We assume instantaneous intermediate node error probability knowledge and MAP detection at the receiver. Hence we may directly make use of the minimum distance values for a given generator matrix $\mathbf{G}$ in determining the error performance through diversity orders without considering error propagation. Therefore, in the following sections, we identify the minimum distances for each source symbol in order to characterize the error performance of a network coded system.
\subsection{A Network Code Example}
\label{sec:nc_const}
\begin{figure}[htbp]
\centering
\epsfig{file=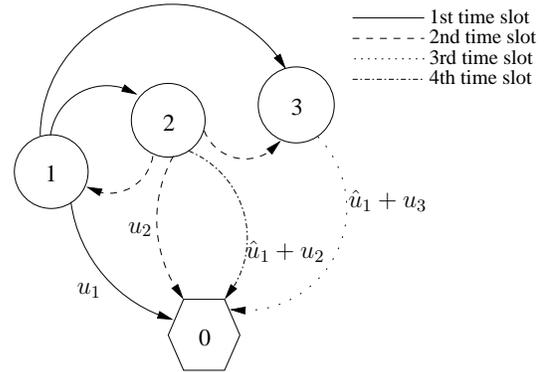,width=0.40\textwidth}
\caption{Sample network coded transmission scenario} 
\label{fig:network3}
\end{figure}

Let us consider an example network code with $n=4$ transmission slots, $k=3$ sources and transmissions over GF($2$) with data rate $r=k/n=\frac{3}{4} \; bits/transmission$:
\begin{align}
\label{eqn:gen}
\mathbf{G}=\left[ \begin{array}{cccc}
1 & 0 & 1 & 1 \\
0 & 1 & 0 & 1 \\
0 & 0 & 1 & 0 \end{array} \right], \mathbf{v} = \left[1\ 2 \ 3\ 2\right].
\end{align}
According to the generator matrix $\mathbf{G}$ and the vector of transmitting nodes $\mathbf{v}$, in the first two time slots (corresponding to the first two columns of $\mathbf{G}$ and the first two entries of $\mathbf{v}$), node $1$ and node $2$ transmit $u_1$ and $u_2$ respectively. In the third time slot, node $3$ tries to encode its own data symbol $u_3$ together with the detection result at the first time slot $\hat{u}_1$ through a simple XOR operation over GF(2). In the last slot, once again node $2$ uses the channel to transmit the network encoded data $c_4 =\hat{u}_1 + u_2$ with its own estimate of $u_1$. This single round of network coded transmissions is summarized in Fig.~\ref{fig:network3}. One can show that the minimum distance for $\mathbf{G}$ is $1$. However, we will see that an error event requires at least $2$ bit errors for detection of $u_1$ at receiver node $0$. 

Let all the data bits be equal to $0$ without loss of generality, i.e., $\mathbf{u}=\left[ 0\;0\;0\right]$. Hence the transmitted codeword is expected to be $\mathbf{c}=\left[ 0\;0\;0\;0\right]$ in case of no intermediate node errors. The error event for $u_1$ corresponds to its detection as $1$. This erroneous detection can occur for sequence detections $\hat{\hat{\mathbf{u}}} \in \{[100],[101],[110],[111]\}$. The incorrect codewords $\hat{\mathbf{c}}$ corresponding to these detected vectors are $[1011],[1001],[1110],[1100]$, respectively. When these codewords are compared to the codeword $[0000]$, it is clear that at least $2$ bit errors are needed to cause an error event. Hence the diversity order for $u_1$ in this setting is said to be $2$. The erroneous detection for other bits can be investigated in a similar fashion. Focusing on $u_3$ and hypothesizing ${\mathbf{u}}=[000]$, $u_3$ is incorrectly detected when $\hat{\hat{\mathbf{u}}} \in \{[001],[011],[101],[111]\}$. The corresponding codewords are $[0010],[0111],[1001],[1100]$. Therefore, a single bit error can cause erroneous detection of $u_3$. As seen in the example, the error codewords for different data bits are different and may need different number of observation bit errors, which suggests that the error performance for a particular data symbol may differ from that for another. This claim is verified through simulations in Section \ref{sec:num_results}.
\subsection{Greedy Codes}
\label{sec:greedy}
In this study, we make use of linear block codes while constructing network codes. 
In comparison with the network coded operation, a repetition coding scheme is also considered. With this scheme, each source node simply transmits its own data in its turn, with no combining operation over GF($M$). Following the $n$ transmissions of $k$ source nodes, the receiver node combines the data received for each source symbol optimally to generate detection results. On the other hand, with network coded operation, we rely on the family of block codes known as greedy codes. These ($n$, $k$, $d$) codes are selected with the following parameters: blocklength (number of transmission slots) $n$, dimension (number of unicast pairs) $k$, and minimum distance (minimum diversity order) $d$. Greedy codes are known to satisfy or be very close to optimal dimensions for all blocklength-minimum distance pairs \cite{Pless93}. Moreover, they are readily available for all dimensions and minimum distances unlike some other optimal codes. 

Let us assume that the network of interest consists of $k=3$ nodes trying to transmit their data symbols over GF($2$). If a round of communication is composed of $n=6$ transmission slots, we deal with codes of type ($6$, $3$, $d$). Starting with the generator matrix and transmitting node vector corresponding to the repetition coding, we have
\begin{align}
\label{eqn:gen2}
\mathbf{G}&=\left[ \begin{array}{cccccc}
1 & 0 & 0 & 1 & 0 & 0 \\
0 & 1 & 0 & 0 & 1 & 0 \\
0 & 0 & 1 & 0 & 0 & 1 \end{array} \right], \mathbf{v}_{\phantom{1}} = \left[1\ 2 \ 3\ 1\ 2 \ 3\right].
\end{align}
It is easily observed that repeating each data bit twice over independent links, this method satisfies only a diversity order of $2$ for all bits $u_1$, $u_2$, and $u_3$. In contrast, using the ($6$, $3$, $3$) greedy code, $\mathbf{G}_1$, a diversity order of $3$ can be the resulting performance figure with the same data rate $1/2$:
\begin{align}
\label{eqn:gen3}
\mathbf{G}_1&=\left[ \begin{array}{cccccc}
1 & 0 & 0 & 1 & 1 & 0 \\
0 & 1 & 0 & 0 & 1 & 1 \\
0 & 0 & 1 & 1 & 0 & 1 \end{array} \right], \mathbf{v}_1 = \left[1\ 2 \ 3\ 1\ 2 \ 3\right].
\end{align}
It should also be noted that greedy codes accommodate each unicast pair with equal diversity order due to the greedy algorithm utilized in their construction. Moreover, contrary to the findings in \cite{Xiao10}, it is easy to obtain any required diversity order for any data bit even using GF($2$). There is no limitation due to number of unicast pairs in terms of desired diversity order. In case we need an increase in data rate and/or have some higher-priority unicast pairs, we can simply omit some columns of a greedy code generator matrix in order to decrease number of transmissions and/or discriminate between pairs. As an example, the following punctured ($5$, $3$, $2$) code is obtained by omitting the last column of $\mathbf{G}_1$ and has a data rate $3/5$ that is higher than those of above two codes:
\begin{align}
\label{eqn:gen4}
\mathbf{G}_2=\left[ \begin{array}{ccccc}
1 & 0 & 0 & 1 & 1  \\
0 & 1 & 0 & 0 & 1  \\
0 & 0 & 1 & 1 & 0  \end{array} \right], \quad\ \mathbf{v}_2 = \left[1\ 2 \ 3\ 1\ 2 \right].\quad
\end{align}
This punctured network code satisfies a diversity order of $3$ for $u_1$; $2$ for $u_2$ and $u_3$. If $u_1$ is of higher priority, this unequal error protection would be preferable especially when the higher rate of the code is considered. As a result, one can choose a network code satisfying desired error protection properties for a determined network size with adequate data rate quite flexibly.

\section{Sum-Product Network Decoder}
\label{sec:spnd}
It is clear that the optimal rule for detection of any unicast transmission symbol $u_i$ grow exponentially in complexity, since (\ref{eqn:opt_det}) requires additions and multiplications growing exponentially in number of users $k$ and/or number of transmissions $n$. Therefore, this rule becomes quickly inapplicable even for moderate-size networks. Recently the sum-product iterative decoder, which is often utilized for decoding of low-density parity-check (LDPC) codes and is a linear-time algorithm, is suggested for decoding general linear block codes as well. In \cite{Moon04}, an idea on the performance of sum-product decoding of block codes with emphasis on the weight of the parity check matrix is given. 

\begin{figure}[htbp]
\centering
\epsfig{file=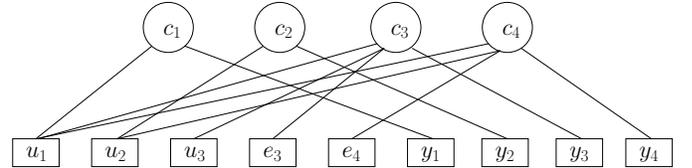,width=0.486\textwidth}
\caption{Tanner graph for network coded system of (\ref{eqn:gen})}
\label{fig:tanner}
\end{figure}

Here, under Rayleigh fading assumption detailed in Section \ref{sec:wn_model_optimal}, we use sum-product decoding and compare its performance with that of the optimal detector. In addition to the variable (the coded symbols, $c_j$) and the check (data symbols and the observations, $u_i$ and $y_j$) nodes describing the linear block code structure of the network code, we should include the check nodes corresponding to the intermediate node errors in the Tanner graph as well. Hence for the network coded system given in (\ref{eqn:gen}), we add two nodes $e_3$ and $e_4$ denoting possible errors at time slots $3$ and $4$. We refer to the graph presented in Fig.~\ref{fig:tanner} for sum-product decoding at the receiver node, namely node $0$. 
It is seen that this simple graph has no cycles in it.

The sum-product algorithm requires the log-likelihood ratios (LLRs) for the check nodes $u_1$, $u_2$, $u_3$, $e_3$, $e_4$, $y_1$, $y_2$, $y_3$, $y_4$ at the initialization step. The data bits ($u_i$'s), which are assumed to be $0$ and $1$ with equal probability, should be simply initialized to LLR values of $0$. The LLRs of error bits are calculated using the related reliability information:
\begin{equation}
\text{LLR(} e_j \text{)} = \ln \frac{1-p_{e_j}}{p_{e_j}},
\end{equation}
where $p_{e_j}$ is the probability that $v_j$ made error(s) in detection of an odd number of data bits that are used in its NC rule $\mathbf{g}_j$. The LLRs for the observations ($y_j$'s) can be calculated using the Gaussian noise distribution (see Section \ref{sec:wn_model_optimal}) as
\begin{equation}
\text{LLR(} y_j \text{)} = \frac{4\mathfrak{Re} \left\lbrace h_j^* y_j \right\rbrace}{N_0},
\end{equation}
where $h_j^*$ is the conjugated gain of the channel over which the modulated symbol $s_j = \mu (\hat{c}_j)$ is transmitted by node $v_j$. Following the initialization step, the sum-product decoder carries on iterations over the given Tanner graph to generate the estimated a posteriori LLRs for the data bits. The number of iterations used and other operational parameters for the decoder are given in Section \ref{sec:perf_sum}.
\section{Numerical Results}
\label{sec:num_results}
\subsection{Sample Network-I: Simulation Results}
\label{sec:num_results_sub1} 
The results in this subsection are based on Sample Network-I of (\ref{eqn:gen}), consisting of only $4$ nodes in order to observe the fundamental issues. At least $100$ bit errors for each data bit $u_1$, $u_2$, and $u_3$ are collected through Monte Carlo simulations for each SNR value. In each run, data bits, intermediate node errors and complex channel gains are randomly generated with their corresponding probability distributions. The solid lines show the BER values for the optimal detector operating under the realistic scenario of intermediate node errors, whereas the dashed lines depict the performance of the genie-aided no-intermediate-error network with the same optimal detection. Finally, the dotted lines are for the detector that neglects possible intermediate errors. 
 
To start with, different diversity orders for bits of different nodes are apparent for optimal detection under intermediate errors. The diversity order for $u_1$ may be observed to be $2$ according to the slope of the corresponding BER curve. This is also given in the Section \ref{sec:nc_const} such that an error event corresponds to at least $2$ bit errors for the detection of $u_1$ and $u_2$. It is clear that no loss of diversity occurs, only an SNR loss of $1.5$ dB for $u_1$ and $u_2$ is evident with respect to the no-intermediate-error operation. Hence the optimal detection rule of (\ref{eqn:opt_det}) is said to avoid the problem of error propagation. The loss for $u_3$, whose diversity order is $1$, with respect to the hypothetical no-intermediate-error network is around $2.5$ dB. The performance deteriorates significantly for especially $u_1$ and $u_2$ when intermediate errors are neglected in detection (dotted lines), i.e., $p_{e_3}=p_{e_4}=0$ is assumed. Not only an SNR loss is endured but also the diversity gains for them disappear.
\begin{figure}[htbp]
\centering
\epsfig{file=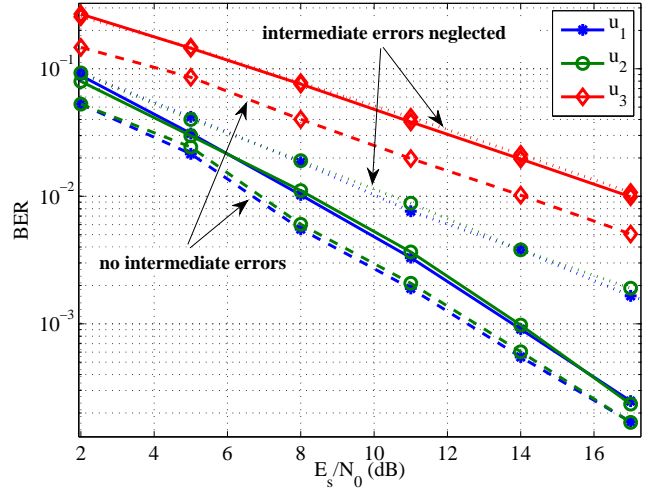,width=0.488\textwidth}
\caption{BER curves for data bits of different nodes for optimal detection}
\label{fig:ber_optimum}
\end{figure}
\subsection{Sample Network-II: Simulation Results}
\label{sec:num_results_sub2}
Next, we compare the performance figures of repetition coding (no network (n/w) coding with simple repetition of source symbols) with two scenarios implementing NC through use of linear block code structures. Two sets of NC generator matrices with vectors of transmitting nodes are the ones given in Section \ref{sec:greedy} in (\ref{eqn:gen3}) and (\ref{eqn:gen4}) respectively.

\begin{figure}[htbp]
\centering
\epsfig{file=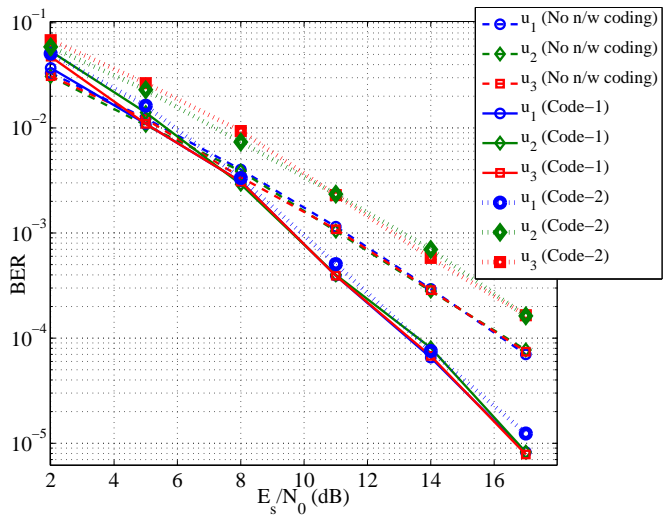,width=0.488\textwidth}
\caption{BER performance for repetition coding and NC with greedy codes}
\label{fig:ber_6_3}
\end{figure}

In fact, the repetition coding method, with $\mathbf{G}$ and $\mathbf{v}$ in (\ref{eqn:gen2}), represents a degenerate case of NC transmitting single bit at each time slot. To construct $\mathbf{G}_1$, we make use of the greedy code with blocklength $n=6$, dimension $k=3$ and minimum distance $d=3$. Fig.~\ref{fig:ber_6_3} exhibits the BER curves for a network of $k=3$ nodes with repetition coding (dashed lines), NC scenarios with Code-1 ($\mathbf{G}_1$) with $n=6$ (solid lines) and Code-2 ($\mathbf{G}_2$) with $n=5$ (dotted lines). The optimal detector of (\ref{eqn:opt_det}) is utilized for this simulation. Clearly, Code-1 has superior performance with a network diversity order (average of all data bits' diversity orders) of $3$. With respect to repetition coding scenario, all data bits observe a $3$ dB SNR improvement for BER = $10^{-4}$. For Code-2, on the other hand, related to the puncturing of a greedy code, bits $u_2$ and $u_3$ observe a diversity order of $2$ while $u_1$ observes an order of $3$. With this unequal protection in mind, the network diversity order for Code-2 is $2.33$, which is higher than that of the repetition coding with order $2$. In addition to improved diversity, Code-2 has also the advantage of increased overall rate due to usage of $5$ slots instead of $6$. It is preferable especially for a network that puts higher priority on $u_1$ than on $u_2$ and $u_3$.
\subsection{Performance of Sum-Product Decoding}
\label{sec:perf_sum}
In this section the performance figures for the sum-product iterative network decoder described in Section \ref{sec:spnd} are given in comparison with the optimal detection rule of (\ref{eqn:opt_det}). The network coded communication system of interest is given in (\ref{eqn:gen3}). The number of iterations for the sum-product type decoder is limited to $4$ with no early termination over parity checks. 

\begin{figure}[htbp]
\centering
\epsfig{file=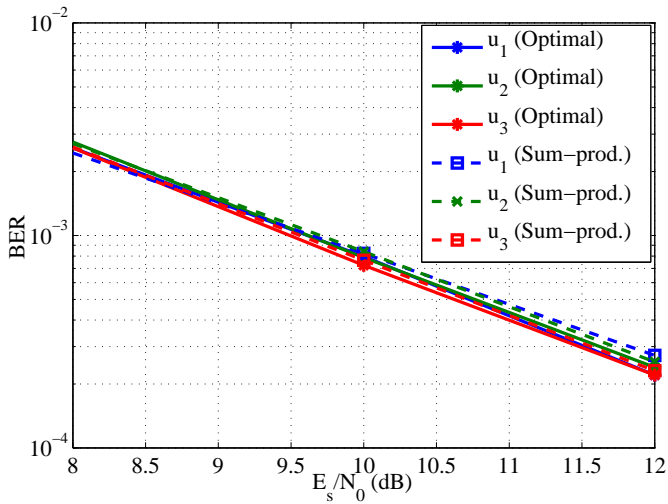,width=0.488\textwidth}
\caption{Optimal decoder of (\ref{eqn:opt_det}) vs. sum-product iterative decoder}
\label{fig:ber_map_sum}
\end{figure}

In Fig.~\ref{fig:ber_map_sum}, we identify the fact that the SNR loss due to usage of sum-product decoder is less than $0.1$ dB for a BER value of $10^{-3}$ for all data bits. Achieving full-diversity with polynomial order of complexity, sum-product type decoding may serve as an ideal method for decoding in network coded wireless systems despite the fact that the corresponding Tanner graph contains cycles. Similar results were also reported previously in \cite{Aktas08, Moon04} for loopy Tanner graphs.
\section{Conclusions}
\label{sec:conc}
Fading and noise in wireless channels exacerbates the problem of cooperative communications in wireless networks. We formulated a network coding (NC) problem for cooperative unicast transmissions. A generator matrix $\mathbf{G}$ is used to represent the combinations performed at relays. The generator matrix and the error probabilities at relays formed the basis for the given MAP-based detection rule. It is found that the performance determining parameter of the scheme depends on the structure of the underlying network code and symbols from distinct source nodes may have different diversity orders. Moreover, the sum-product iterative decoder with polynomial complexity order is shown to perform quite close to the optimal rule. Through our definition of network diversity order, the performance of NC using linear block codes clearly surpasses the repetition coding scenario. Identifying rate and diversity gains of NC for random $\mathbf{G}$ matrices in large networks, studying the effects of imperfect information on channel gains and relay error probabilities, combining channel codes with the described network codes will be addressed in future work. Finally, it would be also interesting to operate suggested wireless NC methods under asymmetrical channel gains.
\bibliography{wcnc_wireless_nc_TA}
\end{document}